\documentclass[preprint,12pt, a4paper]{elsarticle}

\usepackage{amsmath,amssymb,amsfonts}
\usepackage{algorithmic}
\usepackage{graphicx}
\usepackage{textcomp}
\usepackage{xcolor}
\usepackage{balance}
\usepackage{comment}
\usepackage{multirow}
\usepackage[backref]{hyperref} 
\usepackage{makecell}
\usepackage{subcaption}
\usepackage{url}
\usepackage{threeparttable}
\def\BibTeX{{\rm B\kern-.05em{\sc i\kern-.025em b}\kern-.08em
    T\kern-.1667em\lower.7ex\hbox{E}\kern-.125emX}}

\journal{Science of Computer Programming}

\begin{document}

\begin{frontmatter}



\title{CoRaCommit: A VS Code Extension for Commit Message Generation with Exemplar Retrieval}


\author{Chaoran Cai\footnote{Chaoran Cai and Bo Xiong contributed equally to this work.}}
\author{Bo Xiong\footnotemark[1]}
\author{Chong Wang\footnote{Chong Wang and Lulu He are the corresponding authors. Corresponding emails: \texttt{cwang@whu.edu.cn}, \texttt{luluhe@whu.edu.cn}.}}
\author{Lulu He\footnotemark[2]}
\author{Peng Liang}

\address{School of Computer Science, Wuhan University, 430072, China}

\begin{abstract}
Commit messages are essential textual artifacts that describe the intent behind code changes, and play a critical role in version control, code review, and historical tracking. However, in practice, commit messages are primarily authored manually, which is time-consuming and often results in inconsistent quality and non-uniform expression. Existing VS Code extensions for commit message generation typically directly invoke large language models based on the code diff, without leveraging similar commit exemplars as references, and rarely support user feedback-driven LLM recommendation. To address these limitations, this paper proposes CoRaCommit, a VS Code extension that enhances commit message generation by retrieving similar commit exemplars as prompt context, invoking multiple LLMs in parallel for candidate commit message comparison, and dynamically recommending LLMs based on user feedback. Experimental results on 945 commits from the ApacheCM dataset show that CoRaCommit outperforms existing VS Code extensions across BLEU, CIDEr, METEOR, and ROUGE-L metrics, demonstrating the effectiveness of retrieval-augmented context for commit message generation.
\end{abstract}


\end{frontmatter}


\section*{Metadata}
\label{metedata}

Table 1 is the ancillary data table required for the codebase of the extension, and the left column is required to be untouched.

\begin{table}[ht]
\caption{Code metadata (mandatory)}
\centering
\begin{tabular}{|l|p{4.5cm}|p{7.7cm}|}
\hline
\textbf{Nr.} & \textbf{Code metadata description} & \textbf{Please fill in this column} \\
\hline
C1 & Current code version & V2.1 \\
\hline
C2 & Permanent link to code/repository used for this code version &
\makecell[l]{
\href{https://github.com/T-rresa/CoRaCMG}\url{https://github.com/T-rresa/CoRaCMG}
} \\
\hline
C3  & Permanent link to Reproducible Capsule & N/A\\
\hline
C4 & Legal Code License & MIT License \\
\hline
C5 & Code versioning system used &git \\
\hline
C6 & Software code languages, tools, and services used & TypeScript, JavaScript and Python \\
\hline
C7 & Compilation requirements, operating environments and dependencies & VS Code, NPM, Node.js, FastAPI, Docker \\
\hline
C8 & If available, link to developer documentation/manual & N/A \\
\hline
C9 & Support email for questions & charon@whu.edu.cn \\
\hline
\end{tabular}
\label{table2}
\end{table}

\section{Introduction}
Commit messages describe the purpose and content of code changes and play a critical role in software maintenance and collaborative development. Clear and well-structured commit messages help developers quickly understand the intention of code modifications, improve the efficiency of code review, and serve as a vital reference for downstream maintenance tasks such as bug localization and version release management. However, in practice, developers often prioritize implementing features and fixing bugs over documentation. Consequently, commit messages are frequently missing or lack sufficient details~\cite{tian2019goodcommit}, reducing the readability of commit messages themselves and the utility of project history for tasks such as bug localization, version release, and historical tracking.

As large language models (LLMs) continue to advance in code understanding and text generation, automatically generating commit messages from code changes has emerged as a practically significant research topic. 
Existing IDE-integrated commit message generation (CMG) extensions have demonstrated the feasibility of generating commit messages directly within development environments, such as \textit{Auto Commit Message}~\footnote{\url{https://github.com/MichaelCurrin/auto-commit-msg}}, \textit{AI Commit}~\footnote{\url{https://github.com/sitoi/ai-commit}}, and \textit{Commit Sage}~\footnote{\url{https://github.com/VizzleTF/CommitSage.git}}. However, most extensions rely solely on the current code diff as generation input and provide little support for assisting users in selecting among multiple available LLMs.

These limitations motivate the design of CoRaCommit, a VS Code CMG extension that augments LLM with retrieved commit exemplars. CoRaCommit adopts the CoRaCMG~\cite{xiong2026coracmg}, a contextual retrieval-augmented framework that uses  a hybrid retriever combining semantic vectors and BM25 to retrieve similar commit exemplars (i.e., historical diff-message pairs) from the ApacheCM commit dataset~\cite{c3gen} and augments the input prompt with these pairs to guide LLM-based CMG. Beyond this core framework, CoRaCommit provides three key capabilities for IDE-based scenarios: (a) concurrent invocation of multiple LLMs to facilitate candidate comparison; (b) asynchronous feedback-driven evaluation that updates both global and exemplar-level LLM scores based on user selection and editing behavior; and (c) dynamic LLM recommendation, including both a global ranking and task-specific suggestions based on retrieved similar commit exemplars. By leveraging external commits and user feedback, CoRaCommit provides developers with higher-quality commit messages that are better aligned with the current code changes. 

The main \textbf{contributions} of this paper are summarized as follows.
\begin{itemize}
    \item We design and implement \textit{CoRaCommit}, a VS Code extension adopting CoRaCMG framework~\cite{xiong2026coracmg} for automated CMG. It incorporates staged code diff extraction, hybrid retrieval of similar commit exemplars from ApacheCM, and retrieval-augmented generation to deliver candidate commit messages within practical development workflows.
    \item An LLM evaluation and recommendation mechanism is designed and implemented in CoRaCommit to facilitate informed LLM selection for CMG tasks. The proposed mechanism updates both global and exemplar-level LLM scores based on candidate generation results, user selections, and finally adopted commit messages, thereby supporting both global LLM ranking display and task-specific LLM recommendation.
    \item Comparative experiments are performed against existing VS Code CMG extensions to evaluate the quality of commit messages generated by CoRaCommit.
\end{itemize}

The remainder of this paper is organized as follows. Section~\ref{sec:background} presents related work. Section~\ref{sec:descripiton} introduces the system architecture and workflow of CoRaCommit. Section~\ref{sec:features} presents the four core features of CoRaCommit. Section~\ref{sec:experiment} presents the results of the experimental evaluation. Section~\ref{sec:conclusions} concludes the paper and outlines the future work.

\section{Background and Related Work}\label{sec:background}

\subsection{Commit Message Generation}
Research on CMG has progressed through neural machine translation approaches, structured code representation methods, and large language model approaches. Early research primarily modeled the mapping between code diffs and commit messages as a sequence-to-sequence translation problem, treating code diffs as source sequences and commit messages as target sequences~\cite{jiang2017nmtcommit}. Subsequent work further introduced copy mechanisms, abstract syntax trees and other structural information~\cite{liu2019pointergenerator,liu2022atom}, and input condensation methods to enhance the LLM's ability to handle function names, variable names, key terms, and core change content~\cite{zhang2025brevity}.

In recent years, LLMs have demonstrated significant advantages in code understanding and natural language generation tasks, propelling CMG into a new phase. Existing studies indicate that LLMs can generally improve the quality of generated commit messages~\cite{xue2024automated}.

To address the insufficient use of context by LLMs, retrieval-augmented generation methods have been introduced into CMG tasks~\cite{lewis2020retrieval}. These methods provide additional context for the generation process by incorporating relevant examples. For CMG tasks, similar commit exemplars from historical commits can offer expression patterns and terminology information related to the current code diff, thereby helping LLMs better understand the intent behind the modification. For example, CoRaCMG proposes a hybrid retrieval strategy that combines semantic and lexical similarity to retrieve relevant commit exemplars from the historical repository~\cite{xiong2026coracmg}. 

\subsection{VS Code Extensions for Commit Message Generation}
Several VS Code extensions now support commit message generation. Table~\ref{tab:commit-tools} compares these extensions with CoRaCommit.
\begin{table}[ht]
\caption{Comparison of CoRaCommit with Existing VS Code Extensions for Commit Message Generation}
\centering
\footnotesize
\begin{tabular}{|p{3.3cm}|p{2.2cm}|p{2cm}|p{2cm}|p{2.3cm}|}
\hline
\multicolumn{1}{|m{3.3cm}|}{\centering \textbf{VS Code Extensions}} 
& \multicolumn{1}{m{2.2cm}|}{\centering \textbf{Auto Commit Message}} 
& \multicolumn{1}{m{2cm}|}{\centering \textbf{AI Commit}} 
& \multicolumn{1}{m{2cm}|}{\centering \textbf{Commit Sage}} 
& \multicolumn{1}{m{2.3cm}|}{\centering \textbf{CoRaCommit}} \\
\hline
\makecell[l]{Generation \\ approach} 
& Template \& Rule & LLM-based & LLM-based & RAG \& LLM-based \\
\hline
\makecell[l]{Requires external \\ LLM API key?} 
& N/A & Yes & Yes & Yes \\
\hline
\makecell[l]{Input source} 
& Changed file list & Git code diff & Git code diff & Git code diff \\
\hline
\makecell[l]{support for \\ multiple LLMs} 
& N/A & Yes & Yes & Yes \\
\hline
\makecell[l]{Feedback-driven \\ LLM recommendation} 
& N/A & No & No & Yes \\
\hline
\end{tabular}
\label{tab:commit-tools}
\end{table}

\textbf{\textit{Auto Commit Message}} primarily relies on the predefined templates and rules to generate commit messages based on file changes. It generates messages in the format \textit{<type>[scope]: <description>}. Specifically, the tool analyzes file changes to determine the TYPE and optional SCOPE, while the DESCRIPTION is generated by predefined rules that describe what has been created, updated, or renamed. \textit{Auto Commit Message} offers straightforward usage and lower implementation costs, but its generation results depend heavily on predefined rules and exhibit limitations in expressive flexibility and semantic adaptability.

\textbf{\textit{AI Commit}} uses LLMs to generate commit messages from the current code diff. It supports invoking different LLMs to generate Conventional Commits-style commit messages based on the current code diff, providing features such as multi-language support and prompt configuration. Compared to rule-based methods, these extensions offer stronger language expression capabilities, but their generation workflows still primarily revolve around the current code diff.

\textbf{\textit{Commit Sage}} is another VS Code extension that uses LLMs to generate commit messages. It supports multiple AI providers, commit formats, and project-level configurations. This extension advances further in configurability and engineering usage patterns, but its core workflow still primarily involves analyzing the current code diff and generating commit messages without leveraging similar commit exemplars as references.

As highlighted above, existing VS Code CMG extensions exhibit two limitations: first, they rely exclusively on the current code diff for generation, without leveraging similar commits as contextual references; second, they lack support for LLM recommendation when multiple LLMs are available. To address these limitations, this paper introduces CoRaCommit, a VS Code extension that incorporates both retrieval-augmented generation and an adaptive LLM evaluation and recommendation mechanism. Specifically, its retrieval component implements the hybrid retrieval strategy proposed by CoRaCMG~\cite{xiong2026coracmg}. In parallel, CoRaCommit leverages external commit exemplars and historical LLM performance feedback to jointly improve the quality of generated commits and support adaptive LLM selection.

\section{Overview}\label{sec:descripiton}
CoRaCommit is designed to seamlessly integrate into developers' daily commit workflows within VS Code by providing high-quality candidate commit messages without imposing significant extra effort. This section outlines the overall design of CoRaCommit, including its system architecture and workflow of generating commit messages and feedback-driven evaluation.

\subsection{System Architecture}
\begin{figure*}[ht]
\centerline{\includegraphics[scale=0.4]{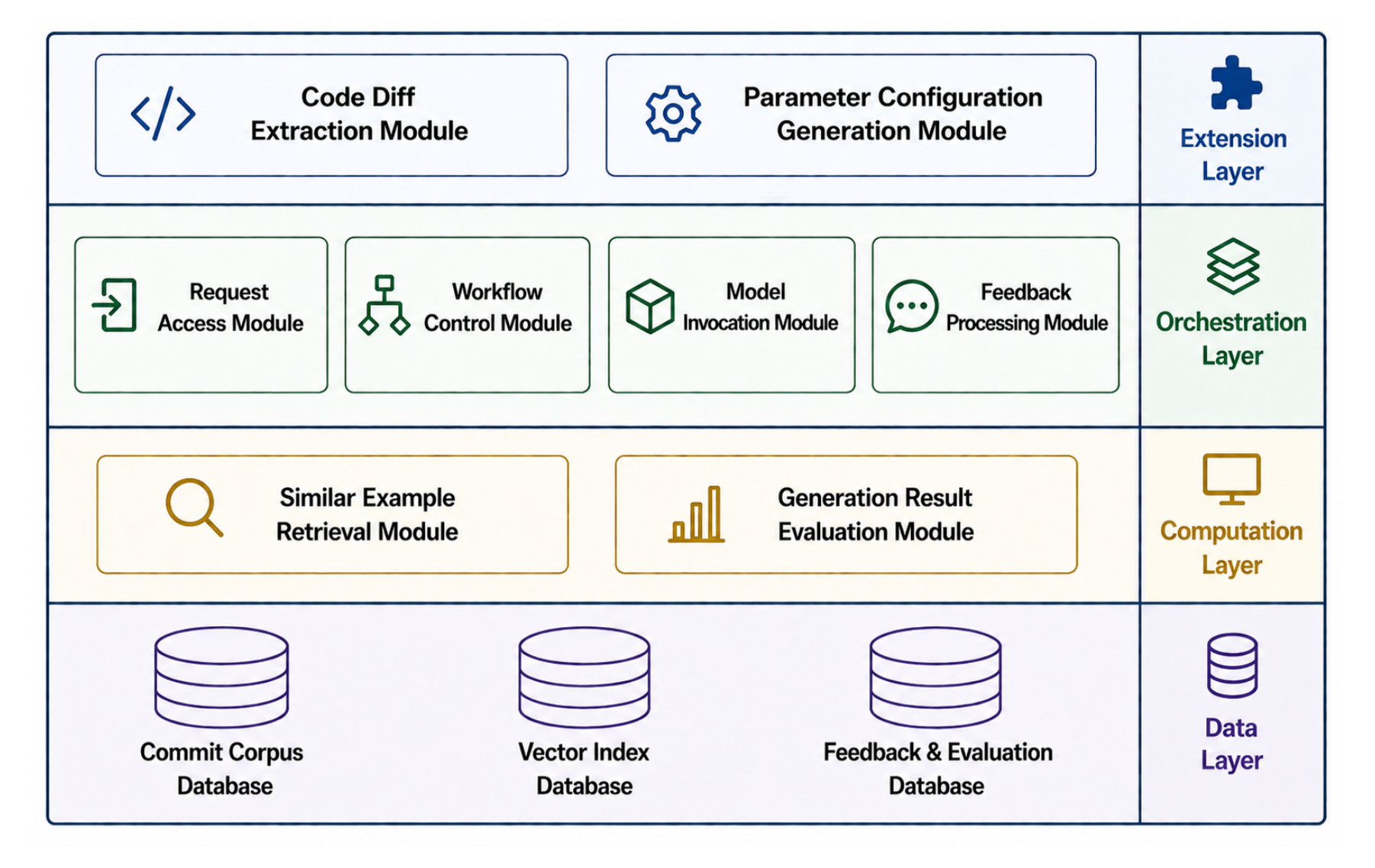}}
\caption{System Architecture of CoRaCommit}
\label{fig_construction}
\end{figure*}

To support interactivity, extensibility, and efficiency in IDE-based workflows, CoRaCommit adopts a layered architecture comprising a VS Code extension frontend, a Node.js service orchestration layer, a Python-based retrieval and evaluation backend, and a foundational data and state management layer. The system architecture of CoRaCommit is shown in Figure~\ref{fig_construction}.

More specifically, the VS Code extension frontend extracts staged code diffs from the current repository, reads user configurations, presents similar commit exemplars and candidate commit messages, and submits feedback after user confirmation. To reduce irrelevant context and mitigate potential privacy risks, CoRaCommit performs lightweight input preprocessing on the client side. The frontend is intentionally designed to remain lightweight and responsive, avoiding complex computation and delegating retrieval, generation, and evaluation tasks to backend services.

The Node.js service orchestration layer handles requests from the extension and coordinates the key processing steps. These steps include retrieving similar commit exemplars, constructing prompts, invoking multiple LLMs concurrently, aggregating results, and enqueueing feedback evaluation tasks for asynchronous processing. In our implementation, this layer is divided into an online API process and a background worker process. The online API process serves extension requests and returns candidate commit messages with low latency, while the background worker consumes feedback evaluation tasks and updates LLM scores asynchronously. Positioned between the extension frontend and the computation services, this layer orchestrates both the online generation pipeline and the asynchronous scoring pipeline.

The Python-based retrieval and evaluation backend retrieves similar commit exemplars and evaluates candidate commit messages generated by the LLMs. It consists of two FastAPI endpoints: Retrieval and Evaluation. The Retrieval endpoint handles vector encoding, candidate retrieval, and reranking, while the Evaluation endpoint performs multi-metric scoring of candidate commit messages.

The data and state layer stores the ApacheCM commit dataset, index files, LLM scores, and user feedback logs. It also maintains queues, caches, and runtime state information to support retrieval lookup, score persistence, queue consumption, and ranking-cache operations.

\subsection{Workflow of CoRaCommit}
\begin{figure*}[ht]
\centerline{\includegraphics[scale=0.48]{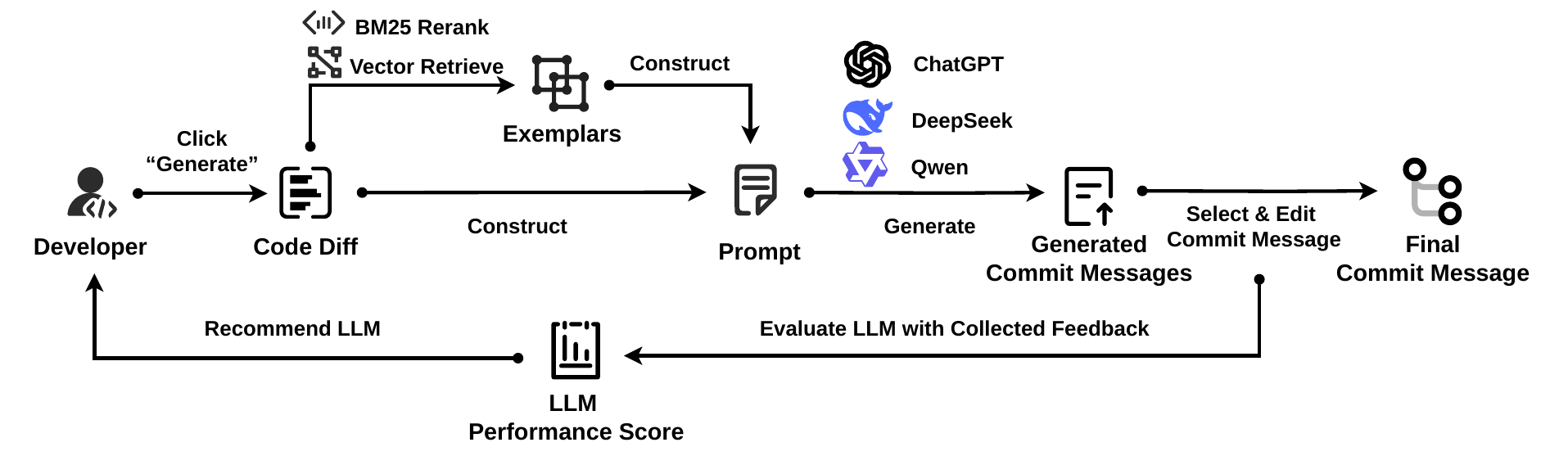}}
\caption{Workflow of CoRaCommit}
\label{fig_workflow}
\end{figure*}

\begin{figure}[ht]
    \centering
    \begin{subfigure}[b]{0.4\textwidth}
        \centering
        \includegraphics[width=\textwidth]{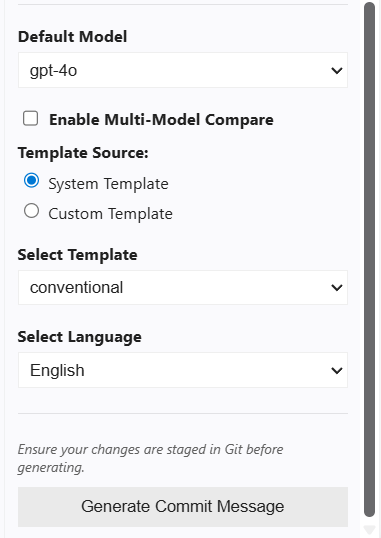}
        \caption{Generation configuration interface.}
        \label{fig-genconfig}
    \end{subfigure}
    \hfill
    \begin{subfigure}[b]{0.4\textwidth}
        \centering
        \includegraphics[width=\textwidth]{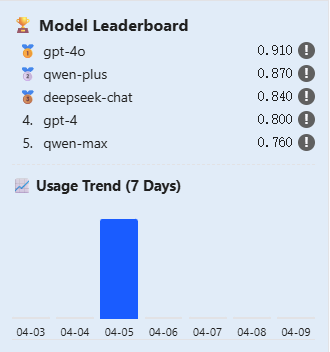}
        \caption{LLM leaderboard interface.}
        \label{fig-LLMrecommend}
    \end{subfigure}
    \caption{Generation configuration and LLM recommendation interface in CoRaCommit.}
    \label{fig-genconfig_recommend}
\end{figure}

\begin{figure*}[ht]
\centerline{\includegraphics[scale=0.7]{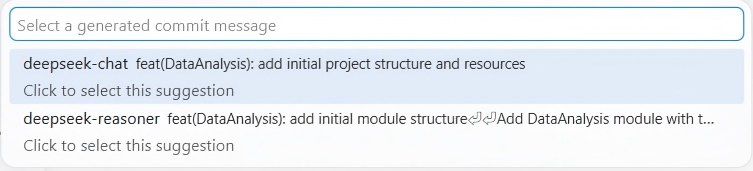}}
\caption{Candidate commit message selection interface.}
\label{fig_can_CM_select}
\end{figure*}

The \textbf{workflow} of CoRaCommit is illustrated in Figure~\ref{fig_workflow}. CoRaCommit begins by extracting the staged code diff from the current repository and performing lightweight local preprocessing. The developer then specifies the target language, the commit message format, and the selected LLMs through the configuration panel shown in Figure~\ref{fig-genconfig}, while the leaderboard interface in Figure~\ref{fig-LLMrecommend} offers supplementary guidance for selecting an appropriate LLM. Based on these settings, the extension sends a generation request to the backend, where the retrieval service searches the ApacheCM dataset for similar commit exemplars. The service orchestration layer combines the processed code diff, retrieved exemplars, and user configurations to construct prompts, and invokes one or more LLMs in parallel to generate candidate commit messages, which are then presented to the developer for selection, as shown in Figure~\ref{fig_can_CM_select}. Once the developer selects or edits a candidate commit message and confirms the commit, CoRaCommit records the interaction as feedback. Subsequently, the background worker process invokes the evaluation service without blocking the online generation pipeline, scores the candidate commit messages, and updates both the global scores of the invoked LLMs and the exemplar-level LLM scores. The updated scores are then reused in subsequent generation tasks, forming a feedback loop among retrieval, generation, evaluation, and recommendation.

Consider a \textbf{representative scenario} where a developer modifies the function \texttt{calculateTotal()} to handle null inputs and stages the corresponding change in Git. CoRaCommit first extracts and preprocesses the staged code diff, and then presents the processed diff for user inspection. Next, the developer specifies the generation settings by selecting English as the target language, \textit{Conventional Commits} as the message format, and three candidate LLMs. After the developer clicks on ``\textit{Generate}'', CoRaCommit retrieves similar commit exemplars such as ``\textit{fix: handle edge cases in payment processing}'', constructs prompts with these exemplars, and generates three candidate commit messages. For example, candidates may include ``\textit{fix: add null check in calculateTotal}'', ``\textit{fix: handle null values in calculateTotal function}'', and ``\textit{refactor: improve null safety in calculateTotal}''. If the developer selects the second candidate and makes minor edits before committing, CoRaCommit records this interaction as feedback and updates the corresponding LLM scores asynchronously.

\section{Core Features of CoRaCommit}\label{sec:features}
This section details the four core features of CoRaCommit: staged code diff extraction, hybrid retrieval-augmented generation, concurrent invocation of multiple LLMs, and dynamic LLM evaluation and recommendation. 
In this section, our VS Code extension is referred to as CoRaCommit.

\subsection{Staged Code Diff Extraction}
To ensure high-quality of subsequent retrieval and generation, the extension first preprocesses the Git staged diff on the client side within the VS Code extension before sending it to backend services. This preprocessing aims to minimize input noise and mitigate privacy leakage risks.

\begin{figure}[htbp]
    \centering
    \begin{subfigure}[b]{0.4\textwidth}
        \centering
        \includegraphics[width=\textwidth]{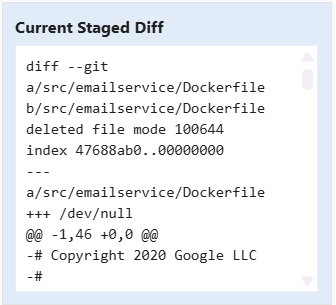}
        \caption{Processed code diff}
        \label{fig-processed_diff}
    \end{subfigure}
    \hfill
    \begin{subfigure}[b]{0.4\textwidth}
        \centering
        \includegraphics[width=\textwidth]{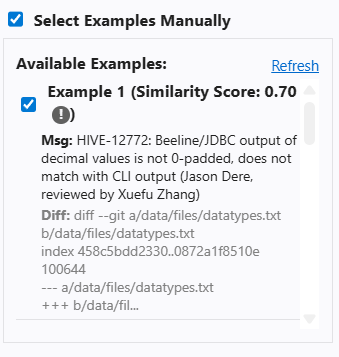}
        \caption{Retrieved similar commit exemplars in sidebar}
        \label{fig-similar_exemplars}
    \end{subfigure}
    \caption{Processed code diff and retrieved similar commit exemplars in sidebar}
    \label{fig-diff_exemplar}
\end{figure}

The process begins by obtaining the code diff content from the staged area of the current repository using \textit{git diff --cached}. The extension then applies preprocessing operations including line ending normalization to handle cross-platform encoding inconsistencies, binary patch filtering to remove non-textual content, sensitive information masking for API tokens and credentials, and length control to respect context window limitations. By performing these steps on the client side, the extension ensures that the code diff content delivered to downstream components is clean, safe, and appropriately sized. As shown in Figure~\ref{fig-diff_exemplar}, the processed code diff is displayed in the sidebar for user inspection.

\subsection{Hybrid Retrieval-Augmented Generation}
To provide LLMs with relevant context beyond the current code diff, the extension adopts a hybrid retrieval approach targeting the ApacheCM commit dataset. The system first encodes the current code diff into dense vector representations using \textit{CodeBERT}~\cite{feng2020codebert} or \textit{jina-embeddings-v2-base-code} embedding models and performs semantic recall based on vector indices to obtain candidate exemplars. Subsequently, these candidates are reranked using BM25 scores to enhance sensitivity to local text features such as function names and path names. This two-stage hybrid retrieval approach balances semantic relevance and keyword matching.

During generation, the top-k similar commit exemplars from ApacheCM are organized into a prompt as reference context for CMG. Similar commit exemplars are also displayed in the sidebar for user inspection (see Figure~\ref{fig-diff_exemplar}).

\subsection{Concurrent Invocation of Multiple LLMs}
The extension supports the parallel invocation of multiple LLMs to generate diverse candidate commit messages, enabling developers to compare alternatives and select the most appropriate one.

\textbf{LLM invocation abstraction.} The Node.js service orchestration layer provides a unified abstraction for LLM invocation, supporting OpenAI-compatible API calls and proxy routing. Each invocation encapsulates an LLM identifier, API credentials, and proxy configuration.

\textbf{Parallel generation.} When multiple LLMs are configured, the orchestration layer concurrently invokes all selected models. This minimizes total waiting time, since the overall latency depends on the slowest response.

\textbf{Result aggregation.} Once all responses have been received, the orchestration layer aggregates the results into a unified format. Each candidate includes the LLM identifier, generated content, and duration of commit generation.

\subsection{Dynamic LLM Evaluation and Recommendation}
\begin{figure*}[ht]
\centerline{\includegraphics[scale=0.80]{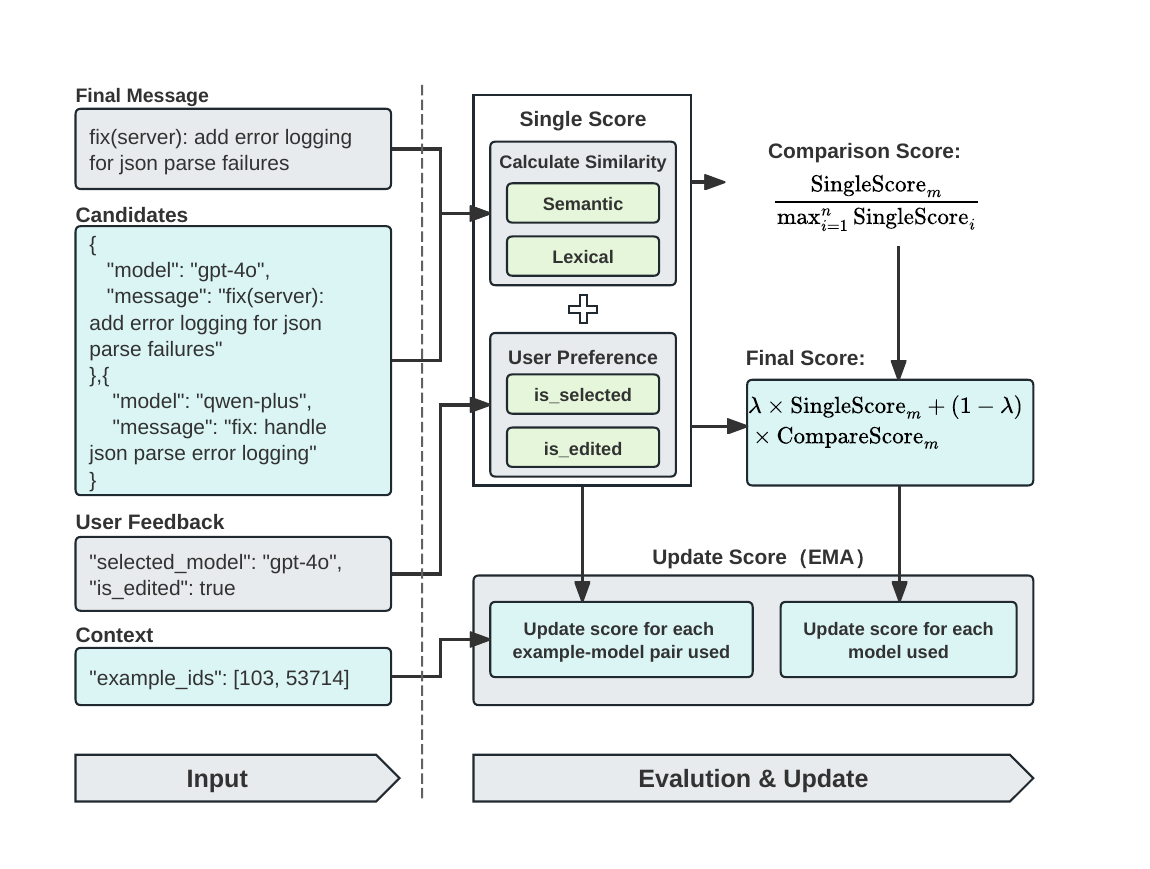}}
\caption{Evaluation and Score Update Process of CoRaCommit}
\label{fig-evaluation-flow}
\end{figure*}

To support more informed LLM selection, the extension provides dynamic LLM evaluation and recommendation. After each commit message is confirmed, the system dynamically updates LLM scores based on user feedback, including candidate selection and editing behavior. This asynchronous update refines LLM scores in near real time without blocking the online generation workflow, as illustrated in Figure~\ref{fig-evaluation-flow}.

For a CMG task involving the candidate LLM set $M$, the system first computes the similarity of the contents between each candidate commit message generated by the LLM and the adopted commit message. The similarity score $\text{SimScore}_m$ is obtained by the weighted combination of semantic and lexical similarity. Additionally, the system constructs a preference score $\text{PreferenceScore}_m$ based on user selection and editing behavior: direct adoption yields a higher score, adoption with edits yields a medium score, and unselected candidates receive a lower score. These metrics are fused into a single LLM score:
\begin{equation}
\text{SingleScore}_m = w_s \times \text{SimScore}_m + w_p \times \text{PreferenceScore}_m
\end{equation}

To mitigate the influence of differences in task complexity, the system normalizes the scores relative to the highest $\text{SingleScore}$ within the same candidate round, yielding $\text{CompareScore}_m$. The final score integrates both metrics:
\begin{equation}
\text{FinalScore}_m = \lambda \times \text{SingleScore}_m + (1 - \lambda) \times \text{CompareScore}_m
\end{equation}

The system then applies exponential moving average to update the global LLM score:
\begin{equation}
\text{GlobalScore}_m^{t} = \gamma \times \text{FinalScore}_m^{t} + (1 - \gamma) \times \text{GlobalScore}_m^{t-1}
\end{equation}

where $\gamma$ denotes the update coefficient. EMA preserves historical information while weighting recent feedback more heavily, making it suitable for scenarios where LLM versions and task distributions evolve over time.

Beyond global scoring, the system also maintains exemplar-level LLM scores that record the performance of each LLM on specific similar commit exemplars. Specifically, \(\mathrm{exemplar}_i\text{-}\mathrm{LLM}_j\) denotes the score of LLM \(j\) on historical generation tasks where  exemplar \(i\) served as the prompt context.  Upon receiving a new generation request, the system first retrieves a set of similar commit exemplars based on the current code diff and computes the similarity score between the code diff of each exemplar and the current code diff. Subsequently, the system retrieves the corresponding exemplar-level LLM scores for these exemplars and performs a weighted aggregation of each LLM's historical performance on this exemplar set, using exemplar similarity as the weight. Based on this aggregation, the system generates a task-specific LLM recommendation.

These four features jointly support the main workflow of the extension. 


\section{Experimental Evaluation}
\label{sec:experiment}
To verify the effectiveness of CoRaCommit in CMG tasks, this paper conducts comparative experiments primarily from the perspective of generation quality, comparing CoRaCommit with existing VS Code CMG extensions on a unified sample set.

The experimental dataset comprises 945 commits randomly sampled from the ApacheCM commit dataset. Each sample in the dataset contains the code diff content and human-written reference commit message. All extensions are evaluated on the same samples with a unified automatic evaluation process for metric calculation. For LLM-based extensions including AI Commit, Commit Sage, and CoRaCommit, the experiments uniformly use \texttt{deepseek-chat} as the external generation LLM, with generation language set to English. CoRaCommit is uniformly configured with \textit{Conventional Commits}~\cite{conventionalcommits} style for all samples, retrieving similar commit exemplars from the ApacheCM commit dataset as prompt context during the generation phase. To ensure a fair evaluation, when retrieving similar commit exemplars for CoRaCommit during the generation phase, the system explicitly excludes the sample itself from the candidate set. 


The experiments employ four metrics, i.e., BLEU~\cite{papineni2002bleu}, CIDEr~\cite{vedantam2015cider}, METEOR~\cite{banerjee2005meteor}, and ROUGE-L~\cite{lin2004rouge}, to measure the quality of generated commit messages in terms of lexical matching, key information coverage, semantic consistency, and overall expression structure. These metrics are widely adopted in natural language generation (NLG) tasks and have been demonstrated to offer complementary perspectives for evaluation. To ensure comparability with prior work, recent studies on CMG have consistently adopted this evaluation protocol~\cite{zhang2024using, xiong2026coracmg}. The corresponding experimental results are presented in Table~\ref{tab:comparative-results}.

\begin{table}[htbp]
\small
\centering
\caption{Experimental Comparison of CoRaCommit with Existing VS Code Extensions for Commit Message Generation}
\label{tab:comparative-results}
\begin{tabular}{l c c c c}
\hline
\textbf{Extension} & \textbf{BLEU} & \textbf{CIDEr} & \textbf{METEOR} & \textbf{ROUGE-L} \\
\hline
Auto Commit Message & 2.13 & 0.036 & 1.31 & 6.37 \\
AI Commit & 1.66 & 0.002 & 17.12 & 11.18 \\
Commit Sage & 2.39 & 0.021 & 19.06 & 14.18 \\
\textbf{CoRaCommit} & \textbf{5.28} & \textbf{0.432} & \textbf{21.40} & \textbf{20.92} \\
\hline
\end{tabular}
\end{table}

From the BLEU and CIDEr metrics, CoRaCommit achieves 5.28 and 0.432 respectively, both higher than the three compared extensions, indicating that its generation results are closer to human-written reference commit messages in terms of local lexical matching and key information coverage. The improvement in CIDEr is particularly notable, suggesting that CoRaCommit better summarizes the core content of code modifications. This result may be attributed to its retrieval of similar commit exemplars from ApacheCM before generation, where the external commits provide the LLM with additional modification expressions and task clues.

On METEOR and ROUGE-L metrics, CoRaCommit achieves 21.40 and 20.92 respectively, also maintaining the best performance. These results indicate that commit messages generated by CoRaCommit not only have advantages in keyword coverage but are also closer to human reference text in semantic consistency, natural language expression, and overall sentence structure. In contrast, although AI Commit and Commit Sage also rely on LLMs and perform better than rule-based extensions on METEOR, their coverage of modification intent and key information remains insufficient due to primarily generating directly from the current code diff.

Overall, Auto Commit Message shows relatively weak overall performance across all four metrics, indicating that while rule-based or template-based methods have lower usage costs, they struggle to adapt to complex and diverse code changes. AI Commit and Commit Sage perform relatively better on semantic relevance metrics, demonstrating the advantages of LLMs in natural language generation. Among these extensions, Commit Sage is the strongest baseline after CoRaCommit, yet it still trails CoRaCommit in BLEU, CIDEr, and ROUGE-L. Overall, the results in Table~\ref{tab:comparative-results} indicate that by introducing similar commit exemplars as retrieval-augmented context, CoRaCommit can generate commit messages more accurately and completely based on code changes.

\section{Conclusions and Future Work}\label{sec:conclusions}
This paper presents CoRaCommit, a retrieval-augmented commit message generation extension for VS Code. CoRaCommit unifies staged diff preprocessing, hybrid retrieval from the ApacheCM commit dataset, concurrent invocation of multiple LLMs, and dynamic LLM evaluation and recommendation within a single workflow. By incorporating similar commit exemplars as retrieval-augmented context, CoRaCommit helps LLMs generate commit messages that better reflect the intent and content of code changes. The comparative evaluation shows that CoRaCommit outperforms existing VS Code CMG extensions on BLEU, CIDEr, METEOR, and ROUGE-L, indicating the effectiveness of exemplar-based context retrieval for improving commit message quality. In addition, the feedback-driven LLM recommendation component supports both global LLM ranking and task-specific recommendations by updating LLM scores based on users' selection and editing behavior.

Despite the promising results, several limitations still exist and leave room for future improvement. First, the effectiveness of retrieval-augmented generation depends on the coverage and quality of the ApacheCM commit dataset, and limited dataset coverage may reduce the usefulness of similar commit exemplars for some code changes. Second, retrieval efficiency requires further optimization when the commit dataset becomes larger or when the system faces higher-concurrency usage scenarios. Third, dynamic LLM evaluation and recommendation require sufficient historical feedback data before global scoring, exemplar-level LLM scoring, and task-specific LLM recommendation can become stable and reliable. Future work will therefore focus on improving retrieval efficiency, expanding the commit exemplar corpus, developing more fine-grained recommendation strategies that better capture task characteristics, and conducting larger-scale user studies to evaluate CoRaCommit's usability and the practical effectiveness of feedback-driven LLM recommendation.



\bibliographystyle{elsarticle-num}
\bibliography{main}


%
%
%

\end{document}